\documentclass[12pt,preprint2]{aastex}   

\def\teq#1{$\, #1\,$}
\def\dover#1#2{\hbox{${{\displaystyle#1 \vphantom{(} }\over{
   \displaystyle #2 \vphantom{(} }}$}}
\def\PitchBnew{\theta_{B}^{N}}
\def\PitchBold{\theta_{B}^{0}}
\def\phiBnew{\phi_{B}^{N}}
\def\phiBold{\phi_{B}^{0}}
\def\rg{r_{g}}

\newcommand{\PAD}{pitch-angle distribution}
\newcommand{\PADs}{pitch-angle distributions}
\newcommand{\DelTheta}{\Delta\Theta_\mathrm{Bn}}
\newcommand{\phiRad}{\phi_{\rg}}
\newcommand{\phiRadold}{\phi_{\rg}^0}
\newcommand{\bEq}{\begin{equation}}
\newcommand{\eEq}{\end{equation}}

\newcommand{\pcc}{cm$^{-3}$}
\newcommand{\Pret}{P_\mathrm{ret}}
\newcommand{\vHT}{v_\mathrm{HT}}
\newcommand{\Delxsk}{\Delta x_\mathrm{sk}}
\newcommand{\delxpf}{\delta x_\mathrm{pf}}
\newcommand{\ucrossB}{{\bf u} $\times$ {\bf B}}
\newcommand{\pparZ}{{p}_{B0}}
\newcommand{\pperpZ}{{p}_{\perp 0}}
\newcommand{\pperpN}{{p}_{\perp N}}
\newcommand{\pZprime}{{p}_{z'}}
\newcommand{\pZZprime}{{p}_{z''}}
\newcommand{\pxold}{p_{x0}}
\newcommand{\pyold}{p_{y0}}
\newcommand{\pzold}{p_{z0}}
\newcommand{\pxtwo}{p_{x2}}
\newcommand{\pytwo}{p_{y2}}
\newcommand{\pztwo}{p_{z2}}
\newcommand{\pxsk}{p_{x,\mathrm{sk}}}
\newcommand{\pysk}{p_{y,\mathrm{sk}}}
\newcommand{\pzsk}{p_{z,\mathrm{sk}}}
\newcommand{\ptsk}{p_{t,\mathrm{sk}}}
\newcommand{\uxZ}{u_{x0}}
\newcommand{\uzZ}{u_{z0}}
\newcommand{\utZ}{u_{0}}
\newcommand{\uttwo}{u_{2}}
\newcommand{\uxtwo}{u_{x2}}
\newcommand{\uztwo}{u_{z2}}
\newcommand{\gamU}{\gamma_{u}}
\newcommand{\gamUold}{\gamma_{u0}}
\newcommand{\gamUtwo}{\gamma_{u2}}

\newcommand{\gampf}{\gamma_\mathrm{pf}}
\newcommand{\gamsk}{\gamma_\mathrm{sk}}
\newcommand{\deltime}{\delta t}
\newcommand{\deltimepf}{\delta t_\mathrm{pf}}
\newcommand{\delsub}{\delta t_\mathrm{sub}}
\newcommand{\Nsub}{N_\mathrm{sub}}
\newcommand{\fofp}{f(p)}
\newcommand{\Bbf}{\mathbf{B}}
\newcommand{\delmax}{\delta \theta_{\rm max}}
\newcommand{\sigTP}{\sigma_\mathrm{TP}}
\newcommand{\uxTwo}{u_{x2}}
\newcommand{\MC}{Monte Carlo}
\newcommand{\etamfp}{\eta_\mathrm{mfp}}

\newcommand{\degg}{^\circ}
\newcommand{\Tbn}{\theta_\mathrm{B0}}
\newcommand{\Tbtwo}{\theta_\mathrm{B2}}
\newcommand{\Tutwo}{\theta_\mathrm{u2}}
\newcommand{\gamZ}{\gamma_0}
\newcommand{\gamray}{$\gamma$-ray}

\newcommand{\TP}{test-particle}
\newcommand{\SC}{self-consistent}

\newcommand{\kmps}{km s$^{-1}$}

\newcommand{\xx}[1]{\times 10^{#1}}

\newcommand{\rel}{relativistic}
\newcommand{\nonrel}{non\-rel\-a\-tiv\-is\-tic}
\newcommand{\transrel}{trans-rel\-a\-tiv\-is\-tic}
\newcommand{\ultrarel}{ul\-tra-rel\-a\-tiv\-is\-tic}

\newcommand\alf{Alfv\'en}

\newcommand\etal{et al.}
\newcommand\itt{ }
\newcommand\bff{ }
\newcount\listnorom
\newcommand\listromanDE{\global\advance \listnorom by 1 
{\lowercase\expandafter{(\romannumeral\listnorom)}\ }}
\newcommand\newlistroman{\listnorom=0}

\newcount\Inum
\newcount\IInum
\newcount\IIInum
\newcount\IVnum
\def\I{\global\multiply\IInum by 0 \global\multiply\IIInum by 0
            \global\multiply\IVnum by 0 \global\advance \Inum by 1
            {\the\Inum. }}
\def\II{\global\multiply\IIInum by 0\global\multiply\IVnum by 0
       \global\advance \IInum by 1 {\the\Inum.\the\IInum. }}
\def\III{\global\multiply\IVnum by 0\global\advance \IIInum by 1
            {\the\Inum.\the\IInum.\the\IIInum. }}
\def\IV{\global\advance \IVnum by 1
            {\the\IVnum. }}
\shorttitle{Relativistic Shock Acceleration}
\shortauthors{Ellison \& Double}

\begin{document}

\title{Diffusive Shock Acceleration in Unmodified Relativistic,
Oblique Shocks}
\author{Donald C. Ellison and Glen P. Double}
\affil{Department of Physics, North Carolina State
 University, Box 8202, Raleigh NC 27695, U.S.A.}
              \email{don\_ellison@ncsu.edu, gpdouble@unity.ncsu.edu}
\email{\rm Accepted in Astropart. Phys., August 2004}

\begin{abstract}
We present results from a fully \rel\ Monte Carlo simulation of
diffusive shock acceleration (DSA) in unmodified shocks. The computer
code uses a single algorithmic sequence to smoothly span the range
from \nonrel\ speeds to fully \rel\ shocks of arbitrary obliquity,
providing a powerful consistency check. While known results are
obtained for \nonrel\ and \ultrarel\ parallel shocks, new results are
presented for the less explored \transrel\ regime and for oblique,
fully \rel\ shocks. We find, for a wide \transrel\ range extending to
shock Lorentz factors $>30$, that the particle spectrum produced by
DSA varies strongly from the canonical $\fofp \propto p^{-4.23}$
spectrum known to result in \ultrarel\ shocks.
Trans-\rel\ shocks may play an important role in
\gamray\ bursts and other sources and most \rel\ shocks will be highly
oblique.
\end{abstract}


\section{Introduction}
\begin{figure}[!hbtp]              
\epsscale{0.95}
\plotone{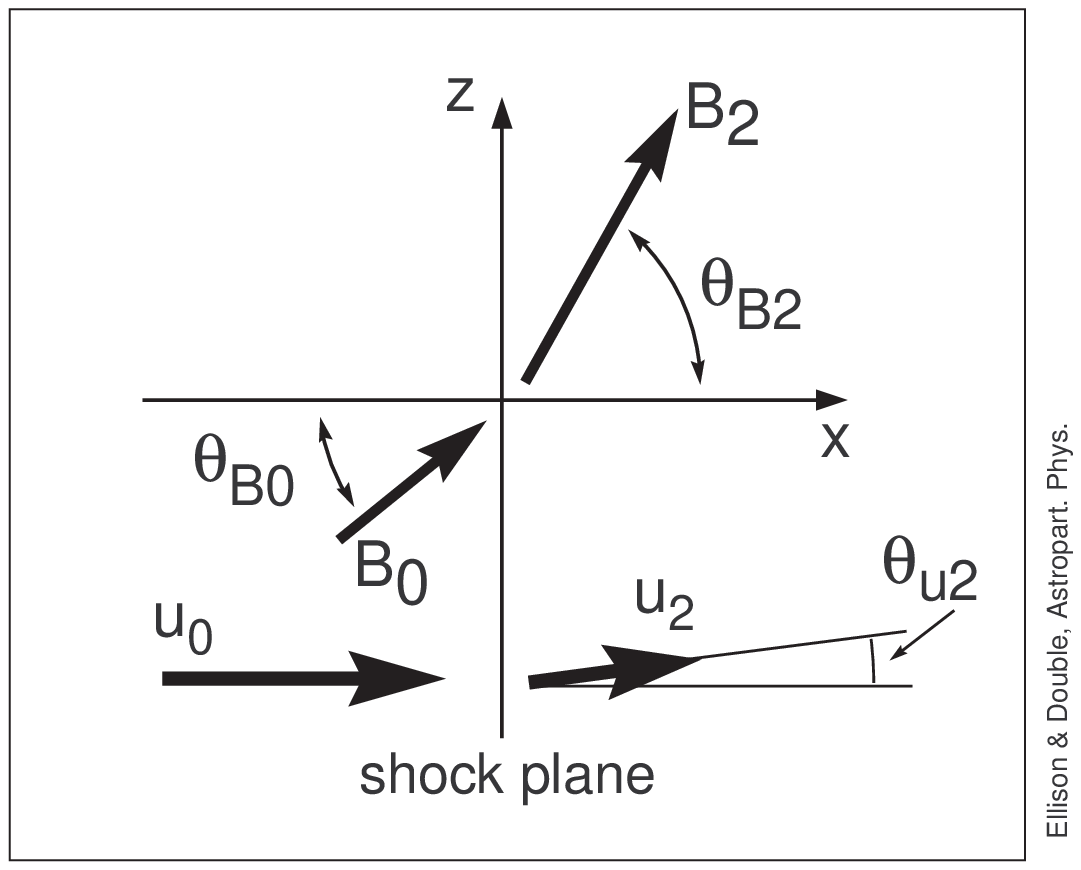}
\figcaption{
Plane shock geometry where the shock lies in the $y$-$z$ plane.
\label{fig_geom}}
\end{figure}

Relativistic shocks, and their associated energetic particles
resulting from diffusive particle acceleration, have received
considerable attention in recent years
\citep[e.g.,][]{Ostrow93,AGKG2001,ED2002,MQ2003a,NO2004}.
This is due primarily to their likely presence in extreme space
phenomena, in particular \gamray\ bursts (GRBs)
\citep[e.g.,][]{Piran99, Rees00}, and their possible role in producing
ultra-high-energy cosmic rays (UHECRs) \citep[e.g.,][]{Waxman2000}.
While most work on diffusive shock acceleration (DSA) in \rel\ shocks
has been restricted to the \ultrarel\ regime with magnetic fields
assumed parallel to the shock normal, \transrel\ shocks are certain to
be important in some sources and, in general, \rel\ shocks will have
highly oblique magnetic fields.
Here we consider \TP\ diffusive shock acceleration in shocks of
arbitrary obliquity\footnote{Oblique shocks are those where the angle
between the upstream magnetic field and the shock normal, $\Tbn$, is
greater than $0\degg$. Parallel shocks are those with
$\Tbn=0\degg$. See Fig.~\ref{fig_geom} for the shock
geometry. Everywhere in this paper we use the subscript 0 (2) to
indicate upstream (downstream) quantities.}
and with arbitrary Lorentz factors, $\gamZ$, ranging from
\nonrel\ shocks to fully \rel\ ones.

Diffusive shock acceleration in relativistic, oblique shocks was
initially addressed by \citet{KH89} and \citet{BH91} both analytically
and numerically, and by \citet{Ostrow91}, using numerical techniques.
\citet{Ostrow93} further investigated oblique, relativistic shocks
using small perturbations superimposed on a uniform magnetic field,
and \citet{BedOstrow96} investigated the corresponding time-scale for
test-particle acceleration.  \citet{BedOstrow98} determined 
the energy spectra of accelerated test particles in oblique,
\ultrarel\ shocks with results leading to a limiting energy
spectral index of 
$\sigma_{E} = 2.3 \pm 0.1$,
independent of shock obliquity.
This result had been
anticipated by \citet{HD88} for test-particle acceleration in
\ultrarel, parallel shocks. 
More recent work on oblique, \rel\ shock acceleration has been
presented by \citet{Meli2002}, \citet{MQ2003b} and \citet{NO2004}.

Despite the emphasis on \ultrarel\ shocks in theoretical work,
\transrel\ shocks may play an important role in GRBs, both for
internal shocks and for the expanding fireball shock presumably
responsible for the afterglow. Even though GRB fireballs may initially
have Lorentz factors $\gamZ > 100$, the internal shocks, credited by
many as responsible for converting the bulk kinetic energy of
expansion into particle internal energy and hence to radiation, may be
much slower with $\gamZ$'s of a few. As the fireball expands into the
ambient interstellar medium, regardless of its initial $\gamZ$, it
will slow and pass through a \transrel\ phase before becoming \nonrel.
Existing afterglow observations span this phase and
understanding Fermi acceleration in \transrel\ shocks is important for
interpreting the observed emission.

Considering the importance of oblique, 
\transrel\ shocks, we have
extended our well-tested \MC\ model of DSA
\citep[e.g.,][]{EJR90,JE91,ED2002} to include shocks of arbitrary
obliquity and speed. We describe the details of the model and present
results for shocks with parameters not previously addressed in
published work. In the two extreme cases where direct comparisons can
be made -- \nonrel\ and \ultrarel\ shocks -- our results are consistent
with previous work.
In future studies we will investigate nonlinear effects in oblique
\rel\ shocks where accelerated particles modify the shock structure
\citep[e.g.,][]{ED2002}. Here we consider only particle
acceleration in plane, unmodified shocks where effects on the shock
structure from superthermal particles are ignored.

In contrast to \nonrel\ shocks, the details of particle scattering
strongly influence the superthermal particle populations produced in
\transrel\ and \ultrarel\ shocks
\citep[e.g.,][]{BedOstrow96,BedOstrow98}. Unfortunately, these details
are not known with any reliability so the results of all current
models of DSA depend on the particular scattering assumptions made.
We use a simple, parameterized model of particle diffusion and attempt
to describe our procedures in sufficient detail so readers can clearly
see how the assumptions influence the particle distribution
functions. While we make no claim that our scattering scheme is more
realistic than other models, some of which are far more complex
\citep[e.g.,][]{NO2004}, we do believe it adequately parameterizes the
particle transport. Until a self-consistent theory of wave-particle
interactions in \rel\ shocks is produced, parameterization will be
necessary.

Unlike most other models of DSA, we follow particles from thermal
energies through the injection process to superthermal energies. While
not necessary in \TP\ calculations, some description of the injection
process from thermal energies must be used in \SC, nonlinear
calculations. The procedures we use to model injection in \rel\
shocks are identical to those we have used with some success in
\nonrel\ shocks \citep[e.g.,][]{EMP90,JE91,BOEF97}.

Finally, while our results are applicable to ion acceleration in GRBs
and the possible production of UHECRs, they are not directly
applicable to the photon emission for two basic reasons. The first is
that energy budget requirements of GRB models generally require
extremely efficient conversion of the bulk kinetic energy of the
fireball into radiation. This means that the particle acceleration
process must be efficient and, therefore, self-consistent, nonlinear
models must be used instead of the \TP\ ones we present here.

Second, the radiation seen from GRBs is produced by electrons not
protons. Normal non-relativistic DSA in a proton-electron plasma will
always put more energy into protons than electrons. In \rel\ shocks,
the fraction of energy going into electrons is reduced
dramatically. In order to determine the distribution of energy between
protons and electrons for GRBs, a two-component acceleration
model must be done in the nonlinear regime. This requires additional
assumptions for the electron injection and, to obtain results even
remotely consistent with GRB properties, requires the modeling of
lepton dominated plasmas. Preliminary work for lepton dominated
parallel shocks is given in \citet{DoublePhd}.

\section{Monte Carlo Simulation of DSA}
Many of the details of the model we use here have been described
previously in the context of \nonrel, oblique shocks
\citep[e.g.,][]{EBJ96,EJB99}, or \rel, parallel shocks
\citep{ED2002}. We refer to this previous work and describe in greater
detail those aspects that are being presented for the first time.

\subsection{Jump conditions}
We let $\utZ$ be the flow speed of the unshocked (i.e., upstream)
plasma as measured in a frame at rest with the shock.  For
convenience, we assume the unshocked flow is normal to the shock along
the $x$-direction so that $\utZ=\uxZ$. 
The shock Lorentz factor is then $\gamma_0 = [1 - (\utZ/c)^2]^{-1/2}$,
where $c$ is the speed of light.  Downstream, the flow will, in
general, have a $z$-component as well, i.e., $\uttwo = \sqrt{\uxtwo^2
+ \uztwo^2}$ (see Fig.~\ref{fig_geom}).
For \nonrel\ shocks, the jump conditions, i.e., the downstream
density $\rho_2$, flow speed $u_2$, pressure $P_2$, and magnetic
field $B_2$, in terms of given upstream values, are well-known.
For \transrel\ shocks, however, the jump conditions are not
straightforward and can only be determined numerically. In all of the
examples given here, except where specifically noted, we assume high
sonic and \alf\ Mach numbers and use the results of \citet{DBJE2004}
to calculate the jump conditions.  

The primary jump conditions come from the equation of continuity which
gives the compression ratio $r = \uxZ/\uxTwo =\gamma_2 \rho_2 /
(\gamma_0 \rho_0)$, and the electromagnetic boundary condition which
determines the angle the downstream magnetic field makes with the
shock normal, $\Tbtwo$. Here, $\gamma_2 = [1 - (u_2/c)^2]^{-1/2}$ and
note that only the components of flow speed along the shock normal are
used in the definition of r.

\subsection{Test-particle power law}
For \nonrel\ shock speeds, the \TP\ power law from DSA is well-known
and is given by:
\begin{equation}
\label{eq:phase}
\fofp\, d^3p \propto p^{-\sigTP}\, d^3p
\ ;
\quad
\sigTP = 3r/(r-1)
\ ,
\end{equation}
where $p$ is momentum and $\fofp\, d^3p$ is the rotationally averaged,
isotropic number density of particles in $d^3p$
\citep[][]{Krymsky77,ALS77,BO78,Bell78}.  For high sonic and \alf\
Mach number, \nonrel, unmodified shocks, the compression ratio
$r\simeq 4$ and the \TP\ power-law index $\sigTP$, is also
approximately 4.\footnote{Note that the energy spectral index for
fully \rel\ particles $\sigma_E$, corresponds to a spectral index in
momentum phase space $\sigma_p = \sigma_E + 2$.}
One of the remarkable aspects of \TP\ DSA is that for $v_p \gg \utZ$,
where $v_p$ is the particle speed measured in the local plasma frame,
$\sigTP$ depends only on $r$ regardless of the diffusive properties of
the plasma and eqn.~(\ref{eq:phase}) holds regardless of the shock
obliquity \citep[e.g.,][]{BO78}.
In other words, provided that the particles are nearly isotropic
simultaneously in the upstream, shock, and downstream reference
frames, eqn.~(\ref{eq:phase}) holds.

For \rel\ shocks, $\utZ \sim c$ and particle distributions are
{\it{not}} isotropic across reference frames. This does not change the
physics of the shock acceleration mechanism -- particles still gain
energy by scattering back and forth across the shock -- but it does
make obtaining analytic descriptions of the acceleration process far
more difficult because the diffusion approximation can no longer be
made.
There are two important consequences of this. First, \MC\ computer
simulation techniques, which do not need to make any assumption
concerning the isotropy of the particle distributions, become the
method of choice for studying particle acceleration
\citep[e.g.,][]{KS87,EJR90,Ostrow91,AGKG2001,ED2002,MQ2003a}, and
second, due to the mathematical difficulties, there are no analytic
solutions to compare computer simulation results to as there are with
\nonrel\ shocks.

\subsection{Details of the \MC\ method}
\MC\ techniques, while straightforward in principle, are, in fact,
difficult in practice with many subtle features which can produce
errors if not implemented properly.  To minimize the possibility of
error, we employ a code which has a {\it single algorithmic sequence}
regardless of $\gamma_0$, $\Tbn$, or scattering parameters.
This allows us to smoothly span the parameter space from \nonrel\ flow
speeds, where analytic solutions exist, through the \transrel\ regime
where analytic solutions do not exist, to \ultrarel\ shocks, where
canonical results again exist for parallel shocks. While a smooth
transition between known results is not a sufficient condition to
guarantee accuracy, it is a necessary condition for correct
results.

\begin{figure}[!hbtp]              
\epsscale{1.0}
\plotone{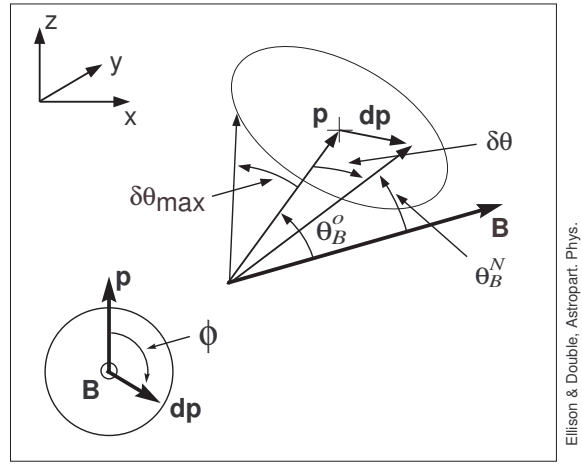}
\figcaption{
Schematic representation of pitch-angle diffusion.
\label{fig_PAD}}
\end{figure}

\begin{figure}[!hbtp]              
\epsscale{1.0} \plotone{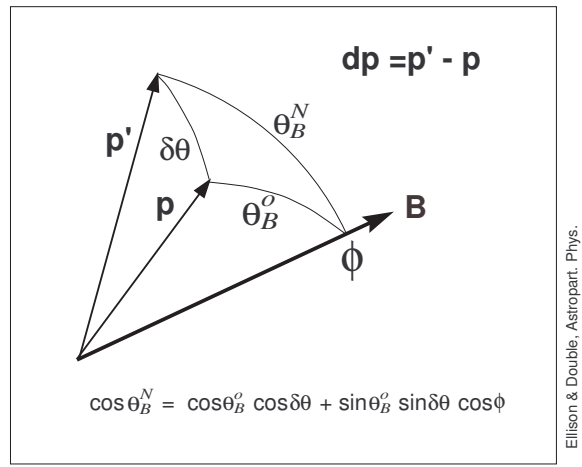} 
\figcaption{ 
Spherical triangle representation of a scattering event where the
momentum changes from {\bf p} to $\mathbf{p'}$ and a particle's pitch
angle changes from $\PitchBold$ to $\PitchBnew$.
\label{fig_sph_tri}}
\end{figure}

\subsubsection{Particle diffusion and convection}
Our implementation of pitch-angle diffusion is described in detail in
\citet{EJR90} and \citet{EBJ96}. We assume that scattering is elastic
and isotropic in the local plasma frame and that the scattering mean
free path $\lambda$ in the local frame is proportional to the
gyroradius $r_g$.  We simulate small-angle scattering by allowing the
tip of the particle's fluid-frame momentum vector $\mathbf{p}$ to
undergo a random walk on the surface of a sphere (see
Fig.~\ref{fig_PAD}).  If the particle originally had a pitch angle
${\PitchBold = \arccos\{ {\bf p}\cdot { \bf B}/\vert {\bf p}\vert
\,\vert{\bf B}\vert\} }$, and after a time ${\delta t}$ undergoes a
small change in direction of magnitude ${\delta\theta}$, its new pitch
angle, ${\PitchBnew}$, is related to the old by
\begin{equation}
   \cos\PitchBnew = \cos\PitchBold \cos{\delta \theta} +
   \sin{\PitchBold} \sin{\delta \theta} \cos{\phi}       
\label{eq:pitchnew}
\ ,
\end{equation}
where \teq{\phi} is the azimuthal angle of the momentum change
$d\mathbf{p}$ measured relative to the plane defined by the original
momentum \teq{{\bf p}} and \teq{{\bf B}} (see
Fig.~\ref{fig_sph_tri}). After each scattering, a new phase angle
around the magnetic field, \teq{\phiBnew}, is determined from the old
phase angle, \teq{\phiBold}, by
\begin{equation}
   \phiBnew = \phiBold + \arcsin \left[ 
   \dover{ \sin{\phi}\, \sin{\delta \theta} }{ \sin{\PitchBnew} } \right] 
  \label{eq:phasenew}
\ ,
\end{equation}
where $\cos{\delta \theta}$ is randomly chosen from a uniform
distribution between 1 and $\cos{\delta \theta_{\rm max}}$, and
\teq{\phi} is randomly chosen from a uniform distribution between
$-\pi$ and $\pi$, so that the tip of the momentum vector walks
randomly over the surface of a sphere of radius \teq{p=\vert {\bf
p}\vert}.

If the time in the local frame required to accumulate deflections of
the order of $90^\circ$ is identified with the collision time $t_c =
\lambda /v_p$, \citet{EJR90} showed that
\begin{equation}
\label{eq:Tmax}
   \delta \theta_{\rm max} =
   \sqrt{ 6 \delta t /t_c }
\ ,
\end{equation}
where $\delta t$ is the time between pitch-angle scatterings.
We take $\lambda$ proportional to the gyroradius $r_g = pc/(eB)$ ($e$
is the electronic charge and $B$ is the local uniform magnetic field
in Gaussian units), i.e., $\lambda = \etamfp \, r_g$, where $\etamfp$
is a measure of the ``strength'' of scattering. The strong scattering
limit, $\etamfp=1$, is called Bohm diffusion.
Now, setting $\deltime = \tau_g / N_g$, where $N_g \gg 1$ is the
number of gyro-time segments $\deltime$, dividing a gyro-period
$\tau_g = 2 \pi r_g/v_p$, we have
\begin{equation}
   \delta \theta_{\rm max} =
   \sqrt{12 \pi / (\etamfp N_g)}
\ ,
\end{equation}
and the scattering properties of the medium are modeled with the two
parameters $\etamfp$ and $N_g$. 

An important limitation of this scheme is the assumption of elastic
scattering which implies that the scattering centers are frozen into
the fluid. This eliminates both the possibility of second-order Fermi
acceleration and the transfer of energy from accelerated particles to
the background plasma via the production and damping
of magnetic turbulence. 
We also neglect any cross-shock electric potential that may exist.

Large $N_g$'s mean particles make many pitch angle scatterings during
a gyro-period, each with a small angular deviation.  The size of the
angular deviation depends on $\etamfp$.  A particular value of
$\etamfp$ means particles will, on average, scatter through $\sim
90^\circ$ while traversing a distance $\sim \etamfp r_g$.  A large
$\etamfp$ implies weak scattering.
This implies, of course, that magnetic fluctuations with sufficient
power exist with correlation lengths on the order of $L_c = 2 \pi
\etamfp r_g / N_g$.
The smaller $N_g$ becomes, the more the scattering resembles
``large-angle scattering'' where the direction of $\mathbf{p}$ is
randomized in a few interactions with the background magnetic field.

For \nonrel\ shocks, $N_g$ has little effect on the results and
then only when $v_p \not \gg \utZ$.
For parallel shocks, regardless of $\gamZ$, $\etamfp$ is unimportant
and the value of $N_g$ required for ``convergence'' increases as
$\gamZ^2$ \citep{ED2002}.  By {\it{convergence}} we mean that as $N_g$
is increased, the spectral index asymptotically approaches a fixed
value.\footnote{When $N_g$ is less than the convergent value, we
  divide the time between scatterings $\deltime = \tau_g / N_g$, into
  subdivisions, $\delsub = \deltime/\Nsub$. In this case, a particle
  is moved without scattering for $\Nsub$ steps and the total
  number of gyro-time segments needed for convergence is $N_g \Nsub$.}

For oblique shocks $\etamfp$ and $N_g$ are both important for the
following reasons.  The size of $\etamfp$ determines the strength of
cross-field diffusion since on average, every time a particle moves
$\sim \lambda$ along $\mathbf{B}$, it will shift $\sim r_g$ across
$\Bbf$. Cross-field diffusion is unimportant in parallel shocks but
plays a critical role in the injection and acceleration processes in
oblique shocks since it makes it easier for downstream particles to
re-cross the shock into the upstream region and be further
accelerated. In fact, with weak cross-field diffusion (i.e., for
$\etamfp \gg 1$), downstream particles will not be able to re-cross an
oblique, \ultrarel\ shock when $\cos{\Tbtwo} \lesssim 1/3$.
With $\Tbtwo$ determined from the jump conditions of
\citet{DBJE2004}, we see that weak
scattering precludes DSA in
essentially all shocks with $\gamZ > 10$ unless they are strictly
parallel.

\begin{figure}[!hbtp]              
\epsscale{0.95}
\plotone{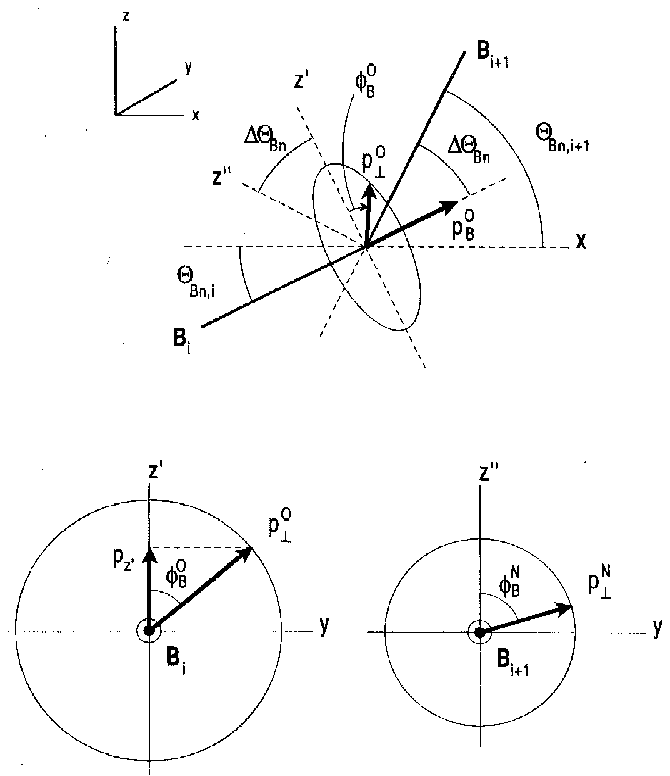}
\figcaption{
Figure 14 from \citet{EBJ96} showing the geometry at the shock where
the magnetic field changes direction and magnitude. In the notation
used here $\Theta_{\mathrm{Bn},i} = \Tbn$, $\Theta_{\mathrm{Bn},i+1} =
\Tbtwo$, $B_i=B_0$, and $B_{i+1} = B_2$.
\label{fig_EBJfig14}}
\end{figure}

We inject particles with a thermal distribution far upstream and allow
them to convect into the shock.\footnote{In all of the results
presented here, the temperature of the unshocked upstream plasma is
low enough that the particle ensemble is \nonrel.}
A particle is translated in the following way.  During the time
$\deltimepf$ between scatterings (measured in the local plasma frame)
a particle will gyrate in the local frame and the local frame will
convect relative to the shock. For the infinite, plane shocks we
consider, only motion in the $x$-direction is important and for motion
in the shock frame we have
\begin{eqnarray}
\label{eq:dummy}  
\Delxsk &=& \gamU [ \delxpf - r_g \sin{\theta_B} 
( \cos{\phiRad} - \nonumber \\
& &  \cos{\phiRadold} ) ] +  u_x \deltimepf
\ ,
\end{eqnarray}
where
\begin{equation}
\delxpf = p_B \deltimepf \cos{\theta_B}/ (m_p \gampf)
\end{equation}
is the $x$-distance the particle moves in the local
frame, 
$p_B$ is the component of momentum along {\bf B} (measured in the
local frame), 
$m_p$ is the mass of a proton, $\gampf=[1 - (v_p/c)^2]^{-1/2}$, $\gamU
= [1 - (u_t/c)^2]^{-1/2}$, where $u_t$ is either $\utZ$ or $\uttwo$,
and $\phiRad$ ($\phiRadold$) is the final (initial) phase of the
gyroradius relative to the $z'$-axis, i.e., in
Fig.~\ref{fig_EBJfig14}, $\phiBold = \phiRadold + \pi/2$.

\subsubsection{Lorentz transformations and shock crossings}
When a particle crosses the shock (where the bulk flow and magnetic
field change direction and magnitude) its pitch angle and phase change
discontinuously.  However, until a scattering occurs, it moves
smoothly and the magnitude of its momentum remains constant in a given
reference frame.  The details of the shock crossing are given in
\citet{EBJ96} and we reproduce Fig.~14 from that paper here as our
Fig.~\ref{fig_EBJfig14}. While the geometry described in \citet{EBJ96}
is identical to that used here, there are important differences in the
techniques we now employ.\footnote{Fig.~\ref{fig_EBJfig14} from
\citet{EBJ96} employs notation suitable for modified shocks where the
field and flow change continuously along the $x$-axis. Since here we
treat only unmodified shocks, the field and flow change only at the
shock at $x=0$.}
The two most important differences are first, that our calculation is
now fully relativistic and applies for all $\gamZ$ and second,
we no longer transform to the de~Hoffmann-Teller (H-T) frame when
crossing the shock. Before discussing the H-T frame, we describe
\citep[following the arguments in][]{EBJ96} the calculations required
in a shock crossing.
For convenience of discussion we describe a particle crossing from
upstream to downstream, although everything we present applies equally
to particles crossing in either direction.

Having determined that a particle has crossed the shock, we first
transform its local frame momentum (in this case, its momentum
measured in the upstream frame, subscript ``0'') to the shock frame:
\begin{eqnarray}
\pxsk &=& 
\left [ (\gamUold - 1) \left ( \frac{\uxZ}{\utZ} \right )^2 + 1
\right ] \pxold + \nonumber \\
& & 
(\gamUold - 1) \frac{\uxZ \uzZ}{\utZ^2} \pzold + \\
& &
\gamUold \gampf  m_p \uxZ 
\ , \nonumber \\
\pysk &=& \pyold
\ , \\
\pzsk &=&
(\gamUold - 1) \left ( \frac{\uxZ \uzZ}{\utZ^2} \right )
\pxold + \nonumber \\
& &
\left [ (\gamUold - 1) \left ( \frac{\uzZ}{\utZ} \right )^2 + 1
  \right ] \pzold + \nonumber \\
& &
\gamUold \gampf  m_p \uzZ
\ .
\end{eqnarray}
Here, $\gamUold = [1 - (\utZ/c)^2]^{-1/2}$. 
Given the shock frame momentum, we now perform a transformation to the
downstream frame. Note that the values of $u_x$, $u_z$, and $\gamU$
change to the downstream values for this transformation:
\begin{eqnarray}
\pxtwo &=&
\left [ (\gamUtwo - 1) \left ( \frac{\uxtwo}{\uttwo}\right )^2
 + 1   \right ] \pxsk + \nonumber \\
& &
(\gamUtwo - 1) \left (\frac{\uxtwo \uztwo}{\uttwo^2} \right ) \pzsk - \\
& &
\gamUtwo \gamsk  m_p \uxtwo
\ , \nonumber \\
\pytwo &=& \pysk
\ , \\
\pztwo &=&
(\gamUtwo - 1) \left ( \frac{\uxtwo \uztwo}{\uttwo^2} \right )
 \pxsk + \nonumber \\
& &
\left [ (\gamUtwo - 1) \left (\frac{\uztwo}{\uttwo} \right )^2 + 1 \right ] 
\pzsk - \nonumber \\
& &
\gamUtwo \gamsk  m_p \uztwo 
\ ,
\end{eqnarray}
where
\begin{equation}
\gamsk = \sqrt{[\ptsk /( m_p c) ]^2 + 1}
\end{equation}
and
\begin{equation}
\ptsk = \sqrt{\pxsk^2 + \pysk^2 + \pzsk^2}
\ .
\end{equation}

Simultaneously with these transformations, the new phase and pitch
angle of the particle relative to the downstream magnetic field must
be computed from the old phase and pitch angle measured against the
upstream magnetic field. This is described in Fig.~\ref{fig_EBJfig14}
where, in our current notation, $\Theta_{\mathrm{Bn},i} = \Tbn$,
$\Theta_{\mathrm{Bn},i+1} = \Tbtwo$, $B_i=B_0$, and $B_{i+1} = B_2$,
since we only consider unmodified shocks.
Referring to Fig.~\ref{fig_EBJfig14}, the component of the old
momentum (i.e., the momentum before crossing the shock) in the
$y$-direction is
\bEq
p_y = \pperpZ \sin{\phiBold}
\ ,
\eEq
and the component along the $z'$-direction is
\bEq
\pZprime = \pperpZ \cos{\phiBold}
\ ,
\eEq
where $\pperpZ$ is the component of momentum (measured in the upstream
frame) perpendicular to $B_0$,
and $z'$ is perpendicular to the
$y$-$B_0$ plane.  The component of momentum along the $z''$-direction
(i.e., the axis perpendicular to the $y$-$B_2$ plane) is:
\bEq
\pZZprime = \pZprime \cos{\DelTheta} - \pparZ \sin{\DelTheta}
\ ,
\eEq
where $\DelTheta = \Tbtwo - \Tbn$.  The total momentum perpendicular
to the new (i.e., downstream) magnetic field $B_2$ is
\bEq
\pperpN = \sqrt{p_y^2 + \pZZprime^2}
\ .
\eEq
Finally, the new phase around $B_2$ is given by
\bEq
\phiBnew = \arctan{\frac{p_y}{\pZZprime}}
\ .
\eEq
The two frame transformations and the pitch angle and phase
transformations occur in the instant the particle crosses the shock
between scatterings.  The fact that we include in the transformations
the changing magnetic field strength and direction and the changing
flow speed and direction, fully takes into account the \ucrossB\
electric field. No transformation to the H-T frame is required.
Furthermore, there is no need to assume the 
conservation of magnetic moment when crossing the
shock.

\subsubsection{de~Hoffmann-Teller frame transformations}
The de~Hoffmann-Teller frame is the frame where the {\bf u} $\times$
{\bf B} electric field is zero. For the geometry shown in
Fig.~\ref{fig_geom}, the H-T frame moves in the negative $z$-direction
with speed $\vHT = \gamZ \utZ \tan{\Tbn}$. As mentioned above, this
frame has been used for oblique shocks, particularly with guiding
center approximations, in attempts to simplify calculations of the
energy change a particle experiences in crossing the shock
\citep[e.g.,][]{EBJ96,MQ2003b}.
Viewed from the shock frame, it appears that particles 
drifting in the shock layer experience an energy change from
the {\bf u} $\times$ {\bf B} field which is not included in normal DSA
where particles gain energy by scattering between the converging
upstream and downstream plasmas.
However, as was evident in the original formulation of \nonrel\ DSA
\citep[e.g.,][]{BO78}, where it was shown that the power-law index was
independent of $\Tbn$, shock drift acceleration must be part of DSA.

From a particle point of view, particles gyrating across an oblique
shock can, depending on their pitch angle, gyrate back and forth
several times before being convected downstream or reflected.
The energy change that occurs in this gyration can be viewed as that
obtained by the particle in the \ucrossB\ electric field as it moves a
distance larger than a gyroradius in the shock layer.
However, since the \ucrossB\ field is zero in the H-T frame and
nothing fundamental changes with a simple frame transformation, the
energy change must be caused by some effect which does not depend
explicitly on the electric field.\footnote{We repeat that our work
ignores any effects from a cross-shock potential which, of course,
cannot be transformed away. We also ignore any other plasma effects
which might occur in the shock layer such as large amplitude waves or
shock surfing.}
In fact, as particles gyrate in the shock layer they continually cross
and re-cross the shock and receive repeated energy changes from the
frame transformations in the converging upstream and downstream
flows. As long as enough scattering occurs to maintain near isotropic
distributions in all frames, this is just the standard DSA process and
gives exactly the same result as obtained from explicitly including
the {\bf u} $\times$ {\bf B} electric field.

For scatter-free propagation, shock-drift acceleration can be viewed
as an independent mechanism. In principle, it is possible, depending
on the pitch angle a particle has in the upstream region, for a
particle to gain a large amount of energy as it gyrates in the shock
layer \citep[e.g.,][]{Decker88}. For some pitch angles, upstream
particles can even gyrate into the shock layer and return upstream,
all without scattering. While in principle it is possible for a
particular particle to gain a large amount of energy in this fashion,
the range in pitch angles that result in large energy gains is
extremely small and only a small fraction of all particles gain
energies much in excess of that from a single shock compression.
Scatter-free propagation, even for \nonrel\ shocks, does not result in
a canonical power law but depends strongly on $\Tbn$ and the energy of
the injected particles.\footnote{For thermal particles in high Mach
number shocks, no particles will be accelerated beyond simple
compression by scatter-free propagation in highly oblique shocks
\citep[e.g.,][]{EBJ95}.}
If elastic
scattering is included, {\it at any level} as long as it is strong
enough to drive the distributions to isotropy, the process reverts to
standard DSA.

In \nonrel\ shocks, a H-T frame can be found unless $\Tbn \simeq
90\degg$ and a H-T transformation may be useful in some
applications. However, a H-T frame with speed $< c$ is excluded when 
$\tan{\Tbn} > c/ (\utZ \gamZ)$,
making this technique essentially useless for \ultrarel, oblique
shocks.

Fortunately, as we have described above \citep[and shown in][]{EJB99},
there is no need to transform to the H-T frame and the effects of the
\ucrossB\ electric field can be included in the particle translation.
This allows us to model \rel\ shocks of arbitrary obliquity.

\subsubsection{Probability of return and distribution function}
The particle spectrum produced by the Fermi mechanism comes about from
the average energy gain per shock crossing combined with the
probability that a particle will make some number of crossings. The
energy gain per crossing is given by the Lorentz transformations
discussed above. The number of crossings a particle makes on average
before escaping from the shock, is determined by the probability that
a particle, once downstream, will be able to scatter back
upstream. This can be determined in the \MC\ code by simply following
a particle as it moves in the downstream region and assuming, once it
reaches some number of diffusion lengths downstream from the shock
that it has a vanishing probability of diffusing back upstream. A more
efficient way of doing this is described in Appendix A3 of
\citet{EBJ96}. Once a particle becomes isotopic in the downstream
frame, the probability of it returning to some point in the downstream
flow is
\bEq
\label{eq:Pret}
\Pret = \left ( \frac{v_p - \uxtwo}{v_p + \uxtwo} \right )^2
\ .
\eEq
This expression is relativistically correct \citep[e.g.,][]{Peacock81}
and applies for any shock obliquity.

For \rel\ shocks, the distribution function $\fofp$ is calculated in
the following way.  As downstream particles leave the shock system,
either from the probability of return test or from a set number of
diffusion lengths downstream, they are binned in momentum space and
$\fofp$ is formed using the shock-frame $\ptsk$.  All of our
distributions with $\gamZ \ge 2$ are calculated this way.

We note that in our previous work \citep[e.g.,][]{JE91,ED2002} we
calculated the distribution at specific locations (i.e., specific
$y-z$ planes) by summing $(\ptsk/|\pxsk|)$ each time a particle
crossed the plane. This produces an omni-directional distribution and
is useful for simulating what a detector will measure at a particular
location relative to the shock. There is an implicit assumption of
isotropy with this technique which can be satisfied for \nonrel\
shocks and, in \nonrel\ shocks, the omni-directional distribution and
the one determined from particles leaving downstream are equivalent as
long as the $y-z$ plane is well downstream and all particles are
isotropic. For \rel\ shocks, anisotropies persist and the two methods
give somewhat different results depending on where the
omni-directional flux is calculated.

\begin{figure}[!hbtp]              
\epsscale{0.95}
\plotone{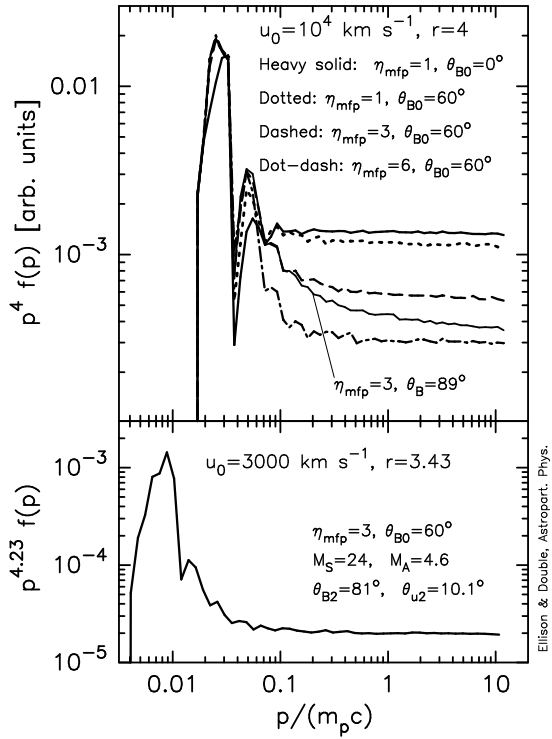}
\figcaption{
Distribution functions, $\fofp$, versus momentum, $p$, for
various input conditions as labeled. Note that we plot $p^4 \fofp$ in
the top panel and $p^{4.23} \fofp$ in the bottom. Spectra with the
expected \nonrel, \TP\ power-law index will be horizontal in this
representation.  The distributions are omni-directional and are
calculated in the shock frame, downstream from the shock.
\label{fig_TP}}
\end{figure}

\newlistroman

\section{Results}
\subsection{Non-relativistic shocks}
In Fig.~\ref{fig_TP} we show $\fofp$ versus $p$ for \nonrel\ shocks
(i.e., $\utZ = 3000$ and $10^4$ \kmps). 
The spectra are omni-directional measured downstream from the shock in
the shock reference frame.
The important features in these plots are:

\listromanDE
In the top panel, the distributions all become power laws with index
$\sigma \simeq \sigTP = 3r/(r-1) \simeq 4$ when $v_p \gg \utZ$. This
confirms that our technique, which does not make a transformation to
the H-T frame, properly accounts for the energy gain in the shock
layer without explicitly including the \ucrossB\ electric field.
To emphasize this point, we show in the bottom panel the spectrum from
a shock with a low \alf\ Mach number $M_A \simeq 4.6$. The parameters
for this shock are: proton density $= 1$ \pcc, unshocked proton
temperature $=10^5$~K, unshocked electron temperature
$=10^6$~K,\footnote{The electron temperature is used only for
determining the Mach numbers.}
and $B_0=3\xx{-4}$~G. In this case, the magnetic field is
strong enough to produce a significant change in downstream flow
direction (i.e., $\Tutwo\simeq 10.1\degg$) and to modify the jump
conditions giving a compression ratio of $r \simeq3.43$. 
Here, we plot $p^{4.23} \fofp$, where $4.23=3 r/(r-1)$, and the
horizontal power law confirms that we get the correct $\sigTP$.
The dynamically important magnetic field makes this a stronger test of
our translation technique than the cases in the top panel where
$B_0=10^{-6}$~G.
For the $\Tbn=60\degg$ shocks in the top panel, the sonic Mach number
$M_S \simeq 81$, $M_A \simeq 4600$ and $\Tutwo \simeq
1.4\xx{-5}$~degrees.

\listromanDE
The normalization of the power law relative to the thermal peak drops
as either $\etamfp$ or $\Tbn$ is increased.  This drop occurs before
particles become isotropic and comes about because particles with 
$v_p \not \gg \utZ$ enter the downstream region directed predominantly
along the shock normal. This makes it harder for them to scatter back
upstream and the difficulty increases when either $\etamfp$ or $\Tbn$
is large.

Particles gyrate around and along the magnetic field. In
the absence of cross-field diffusion, the greater $\Tbtwo$, the
further particles have to move along the field to move a given
distance normal to the shock.  This increases the likelihood that they
will convect downstream without further acceleration. When $\etamfp$
is large, cross-field diffusion is less important compared to
scattering along the magnetic field and particles again are less
likely to scatter back upstream.

Note that the probability of return equation~(\ref{eq:Pret}), which
does not include $\Tbtwo$, is still valid even though particles are
swept downstream more quickly in oblique shocks than in parallel
ones. As a downstream particle gyrates around the magnetic field, it
will cross a particular point in the downstream flow more often if the
field is oblique than if it is parallel. If this point is used to
calculate $\Pret$, the particle will have some probability of escaping
every time it gyrates across this point and is more likely to escape
downstream than would be the case in a parallel shock.  Nevertheless,
the \TP\ power law index $\sigTP$ is obtained once particles become
isotropic since, on average, a particle gains more energy per shock
crossing in an oblique shock than in a parallel one because it can
gyrate across the shock several times in a single shock crossing
event. These two effects, more energy per crossing but smaller
probability of making $N$ crossings, each depend in the same way on
$\Tbtwo$ or $\etamfp$ for isotropic particles and combine to give the
canonical $\sigTP$ once $v_p \gg \utZ$.

\listromanDE
The minimum momentum where the power-law tail obtains increases with
increasing $\Tbn$. This is particularly noticeable with the
$\Tbn=89\degg$ example in the top panel of Fig.~\ref{fig_TP} and comes
about because, to obtain $\sigTP$, particles must have a speed large
compared to the effective flow speed.  In oblique shocks, this speed
is essentially the H-T velocity and the momentum where $\sigTP$ is
established increases with $\Tbn$.

The non-power-law tail coming off the thermal distribution in oblique
shocks may be important in the heliosphere where highly oblique
interplanetary traveling shocks accelerate the thermal solar wind
\citep[e.g.,][]{BOEF97}. The
energetic particles observed by spacecraft may not show the canonical
Fermi power law even though DSA is the physical mechanism producing
the energetic population.

\begin{figure}[!hbtp]              
\epsscale{0.95}
\plotone{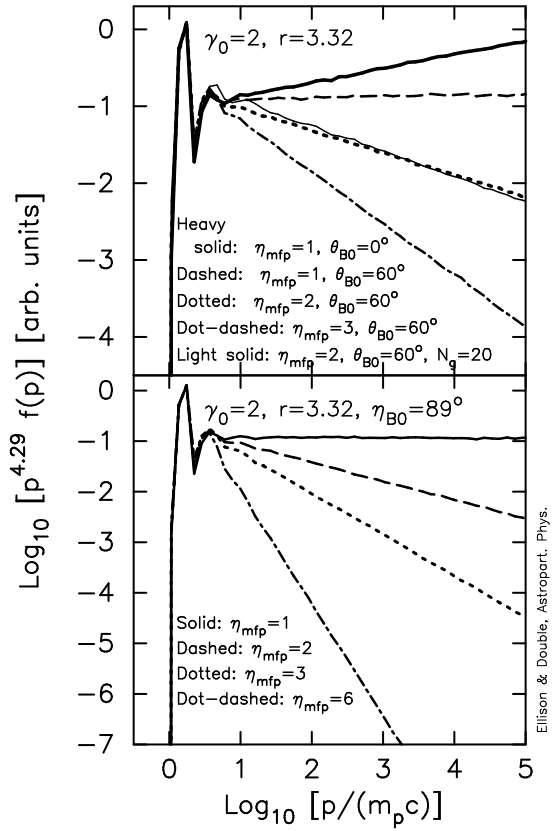}
\figcaption{
Distribution functions, $\fofp$, versus momentum, $p$, for
various input conditions as labeled. Note that we plot $p^{4.29}
\fofp$, where $4.29 \simeq 3 r /(r-1)=\sigTP$, i.e., the spectral index
expected for \nonrel\ shocks with $r=3.32$. 
The light-weight solid curve in the top panel uses a small value of
$N_g$.  These are \transrel\ shocks with $\gamma_0=2$ and show spectra
both flatter and steeper than $\sigTP$.
\label{fig_trans}}
\end{figure}

\subsection{Trans-relativistic shocks}
As soon as the flow speed becomes comparable to the speed of light,
the power-law index starts to depend strongly on $\etamfp$ and $\Tbn$.
In Fig.~\ref{fig_trans} we show results for a \transrel\ shock with
$\gamZ = 2$ for various $\etamfp$ and $\Tbn$. These shocks all have
$r\simeq 3.32$ as determined by the jump conditions given in
\citet{DBJE2004} and we plot $p^{4.29} \fofp$, where $4.29 \simeq 3 r
/(r-1)$. The spectra are measured in the shock frame from escaping
particles.  
In all cases except for the light-weight solid curve in the top panel,
the value of $N_g$ is large enough to produce convergent results.
The heavy-weight solid curve in the top panel is the parallel shock
case and has $\sigma \simeq 4.1$.  The other curves in the top panel
show results with $\Tbn=60\degg$ for various $\etamfp$. A
power law is obtained in each case, but there is a strong steepening
of the power-law portion of the spectrum with increasing $\etamfp$.

In the bottom panel of Fig.~\ref{fig_trans} we show the highly oblique
case $\Tbn=89\degg$ for various $\etamfp$.  The spectra are similar to
those with $\Tbn=60\degg$ except the steepening with increasing $\etamfp$
is greater.

As $\gamZ$ increases, the influence of $N_g$ increases, but it is
still small for $\gamZ=2$. The light-weight solid curve in the top
panel of Fig.~\ref{fig_trans} was calculated with the same parameters
as the dotted curve except that $N_g=20$ rather than 1200 for the
dotted curve. The small $N_g$ produces a slight steepening of the
power law.

\begin{figure}[!hbtp]              
\epsscale{0.95}
\plotone{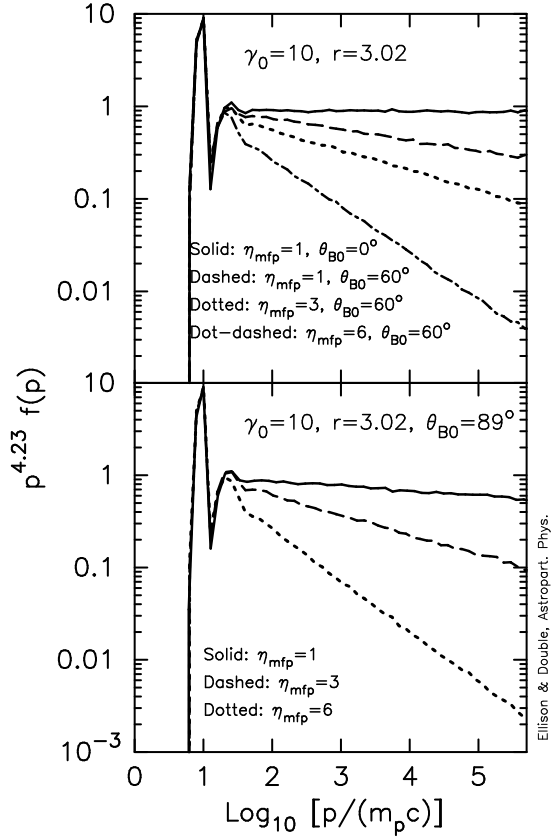}
\figcaption{
Distribution functions, $\fofp$, versus momentum, $p$, for
shocks with $\gamma_0=10$.  Note that the canonical power law index
for \ultrarel\ shocks is $\sigma \simeq 4.23$ and we plot $p^{4.23}
\fofp$ to emphasize this result.
\label{fig_gam10}}
\end{figure}

\begin{figure}[!hbtp]              
\epsscale{0.95}
\plotone{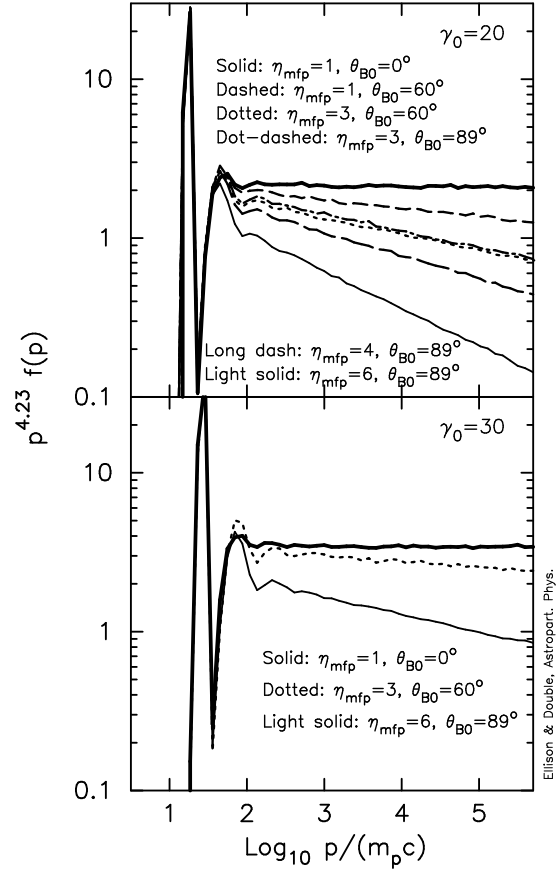}
\figcaption{
Distribution functions, $\fofp$, versus momentum, $p$, for
shocks with $\gamma_0=20$ (top panel) and $\gamZ=30$ (bottom panel).
We plot $p^{4.23} \fofp$ to emphasize the $\Tbn=0\degg$ result.
\label{fig_g20_g30}}
\end{figure}

\subsection{Fully relativistic shocks}
In Fig.~\ref{fig_gam10} we show results with $\gamZ=10$ for various
$\etamfp$ and $\Tbn$.  These shocks all have $r\simeq 3.02$ ($\sigTP
\simeq 4.49$) but we have plotted $p^{4.23} \fofp$, where $4.23$ is
the index known to result for fully \rel, parallel shocks with strong
scattering \citep[e.g.,][]{BH91,BedOstrow98,AGKG2001,ED2002}. In all
cases, the value of $N_g$ is large enough to produce convergent
results.
The solid curve in the top panel confirms that we obtain the canonical
result for $\Tbn=0\degg$.

For $\Tbn > 0\degg$, the results show the same general trend as those
for the \transrel\ shocks in Fig.~\ref{fig_trans}.  The top panels of
Figs.~\ref{fig_trans} and \ref{fig_gam10} are similar except the
steepening for $\gamZ=10$ is a less dramatic function of
$\etamfp$. The same is true for the bottom panels where
$\Tbn=89\degg$.

\begin{figure}[!hbtp]              
\epsscale{0.95}
\plotone{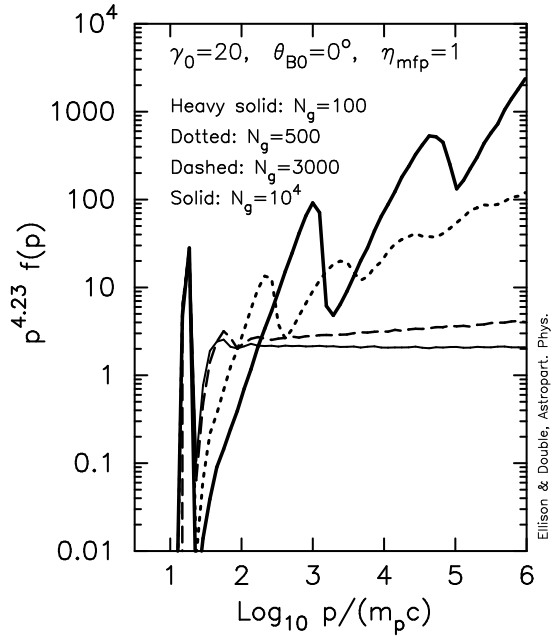}
\figcaption{
Distribution functions, $\fofp$, versus momentum, $p$, for
shocks with $\gamma_0=20$.  The parameter $N_g$ is varied while all
other input parameters are held constant.
\label{fig_g20_vary_Ng}}
\end{figure}

The computation time required to produce convergent results (i.e.,
results with large enough $N_g$) increases with $\gamZ$ so we show a
limited number of examples with $\gamZ > 10$.
Fig.~\ref{fig_g20_g30} shows results for $\gamZ=20$ (top panel) and
$\gamZ=30$ (bottom panel) and these examples, combined with our
$\gamZ=10$ result, show a clear trend.
First of all, the standard $\sigma \sim 4.23$ result for $\Tbn=0\degg$
is obtained in all cases. For $\Tbn>0\degg$, we get results very
similar to those for $\gamZ=10$ with a general reduction in the
variation caused by varying $\Tbn$ or $\etamfp$.
It is likely that this trend will continue, supporting the assertion
that \ultrarel\ shocks produce power laws with $\sigma \simeq 4.23$
independent of $\Tbn$ or $\etamfp$. This assumes, of course, that
$N_g$ is large enough to produce convergence.

The effect of varying $N_g$ when $\gamZ=20$ and $\Tbn=0\degg$ is shown
in Fig.~\ref{fig_g20_vary_Ng} (similar results are obtained for
$\gamZ=10$ and 30). This parameter now dramatically influences the
spectrum.
The step-function effect in the spectrum has been noted by several
authors in the context of large-angle scattering
\citep[e.g.,][]{QL89,EJR90,Baring99} \citep[see
also][]{Vietri95}. Fig.~\ref{fig_g20_vary_Ng} shows that this effect
emerges smoothly from the canonical $\sigma = 4.23$ power law as $N_g$
is lowered and scattering becomes coarser.
Notice that the momentum of the first peak above thermal energies
depends on $N_g$. For $N_g=100$ this peak occurs at $\sim 10^3 m_p c$
rather than $\gamZ^2 m_p c$ because the curves in
Fig.~\ref{fig_g20_vary_Ng} (and all others in this paper) are calculated
in the shock frame. Plotted in the downstream plasma frame, this peak
is at $\sim \gamZ^2 m_p c$ as expected.

\begin{figure}[!hbtp]              
\epsscale{0.95}
\plotone{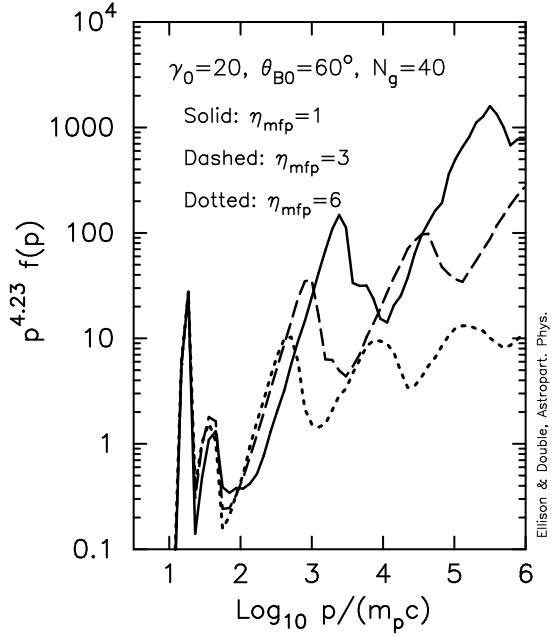}
\figcaption{
Distribution functions with $\gamZ=20$ for small $N_g=200$
and various $\Tbn$ and $\etamfp$, as indicated.
\label{fig_small_Ng}}
\end{figure}

In Fig.~\ref{fig_small_Ng} we show the effect of coarse scattering
(small $N_g=40$) as a function of $\etamfp$ for $\gamZ=20$ and
$\Tbn=60\degg$.  The step-like structure remains as the spectra
steepen with increasing $\etamfp$, although the steps begin to smooth
some.

\begin{figure}[!hbtp]              
\epsscale{0.95}
\plotone{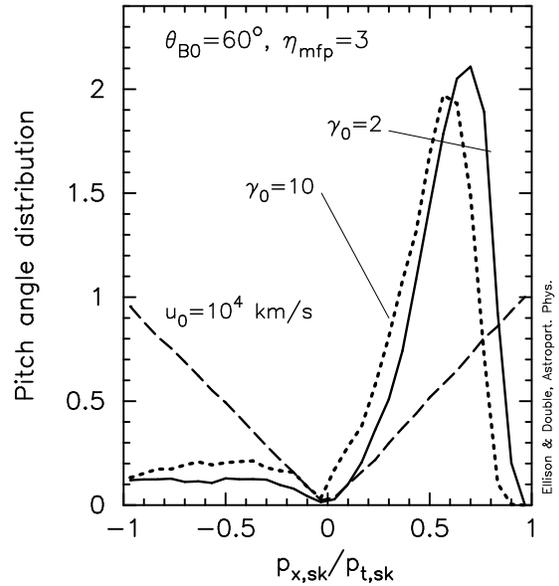}
\figcaption{
Pitch-angle distributions for shock crossing particles in shocks
ranging from \nonrel\ to fully \rel, as indicated.  A positive $\pxsk$
indicates motion toward the downstream direction.  The area under each
curve is normalized to one.
\label{fig_pitch}}
\end{figure}

\subsection{Pitch-angle distributions}
The anisotropic nature of particles in \rel\ shocks produces the
strong dependencies we've seen on $\Tbn$, $\etamfp$, and $N_g$.  The
main difference between \nonrel\ and \rel\ shocks is shown in
Fig.~\ref{fig_pitch}, where we compare the pitch-angle distributions
of shock crossing particles.  The curves are calculated by summing the
quantity $\pxsk/\ptsk$ as particles cross the shock and only
superthermal particles are included in the plots. These are
shock-frame values of momentum and the dashed curve from our \nonrel\
example (Fig.~\ref{fig_TP}) shows the flux weighting characteristic of
an isotropic distribution, i.e., particles cross the shock with a
frequency proportional to their component of velocity normal to the
shock.

The \rel\ shocks show highly anisotropic pitch-angle distributions
with most particles crossing into the downstream region at an oblique
angle. 
For a given set of parameters $\gamZ$, $\Tbn$, $\etamfp$, and
$N_g$, a particular pitch-angle distribution results and this
determines the average momentum gain in a shock crossing. This
combined with the probability to make some number of crossings
determines the spectrum. Unlike for \nonrel\ shocks with their
isotopic distributions, no simple analytic expression has been found
for $\sigma$ once flows become \transrel.

\begin{figure}[!hbtp]              
\epsscale{0.95}
\plotone{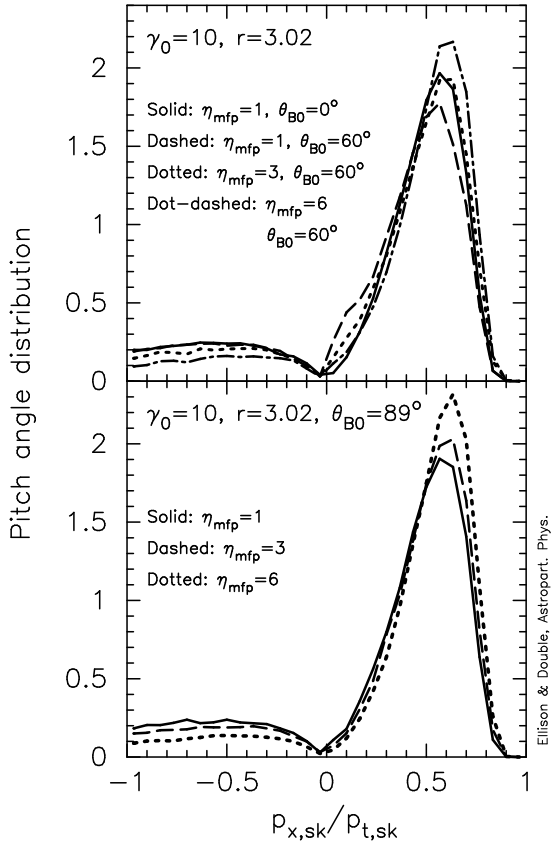}
\figcaption{
Pitch-angle distributions for the shocks shown in
Fig.~\ref{fig_gam10}.  The area under each curve is normalized to one.
\label{fig_pitch_g10}}
\end{figure}

In Fig.~\ref{fig_pitch_g10} we show the \PADs\ for the examples shown
in Fig.~\ref{fig_gam10}.  It's clear that relatively subtle changes in
the \PAD\ causes fairly large changes in $\fofp$. The bottom panel for
$\Tbn=89\degg$ shows the trend as $\etamfp$ increases; the larger
$\etamfp$, the more peaked the distribution is at $\pxsk/\ptsk \sim
0.6$ with a smaller fraction of particles crossing back into the
upstream region.

\begin{figure}[!hbtp]              
\epsscale{0.95}
\plotone{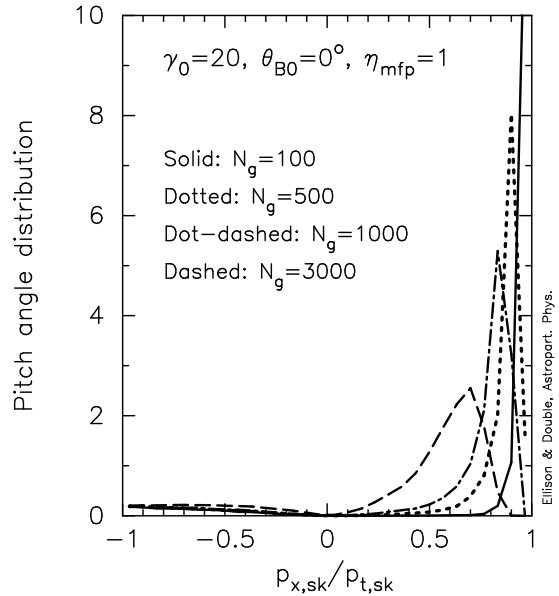}
\figcaption{
Pitch-angle distributions for shocks with parameters similar to those
in Fig.~\ref{fig_g20_vary_Ng}. The area under each curve is normalized
to one.
\label{fig_pitchNg}}
\end{figure}

Finally, in Fig.~\ref{fig_pitchNg} we show \PADs\ as $N_g$ is varied
as in Fig.~\ref{fig_g20_vary_Ng}. When $N_g$ is small, the \PAD\ is
strongly peaked in the forward direction and this is responsible for
the step-like nature of $\fofp$. As $N_g$ is increased, the
distribution smoothly moves to a configuration similar to those shown
in Fig.~\ref{fig_pitch_g10}.

\begin{figure}[!hbtp]              
\epsscale{0.95}
\plotone{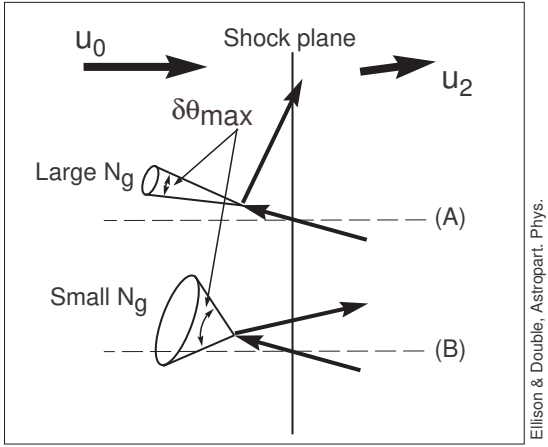}
\figcaption{
Schematic representation of particles crossing a shock with large and
small values of $N_g$.
\label{fig_cross}}
\end{figure}

We illustrate this in Fig.~\ref{fig_cross}.
When $v_p \not \gg \utZ$, flux weighting causes most particles
crossing into the upstream region from downstream to be directed
close to the shock normal, i.e., with pitch angles near $180\degg$.  A
large $N_g$ means a particle will interact after a short $\deltime$
with a small $\delmax$ and is less likely to
broaden its pitch angle much before being swept back across the
shock. This is illustrated in Fig.~\ref{fig_cross} with the particle
trajectory labeled (A). If $N_g$ is small, $\delmax$ is larger, the
particle changes its pitch angle by a larger amount on average, and
there is a greater chance the particle will cross back into the
downstream region with a flatter trajectory as shown with the
trajectory labeled (B). On average, particles crossing as in (A) with
a large $N_g$ receive less energy per crossing than those crossing as
in (B), where $N_g$ is small.

\section{Conclusions}
We have presented a comprehensive study of \TP, diffusive shock
acceleration in plane shocks of varying Lorentz factor $\gamZ$, obliquity
$\Tbn$, and scattering properties parameterized by $\etamfp$ and
$N_g$.  
We use a \MC\ computer simulation which smoothly spans the parameter
space from \nonrel\ shocks through \transrel\ ones to fully \rel\
shocks, without requiring a transformation to the de~Hoffmann-Teller
frame or making any assumptions concerning the magnetic moment of
shock crossing particles. Due to the Lorentz transformation of the
magnetic field, \rel\ shocks will have superluminal H-T speeds for
virtually all obliquities, making transformations to the H-T frame of
limited use.

For \nonrel\ shock speeds, the \TP\ Fermi power law has a well-known
analytic form which is used to test our computer simulations. For
\transrel\ and fully \rel\ shocks, however, analytic results are rare
or non-existent although several independent calculations provide a
canonical power-law for \ultrarel\ particles in parallel, \ultrarel\
shocks, i.e., $\fofp \propto p^{-4.23\pm 0.1}$.  We obtain
this result and the fact that our simulation has a single algorithmic
sequence spanning all $\gamZ$'s and $\Tbn$'s provides some measure of
assurance that our results are accurate in \transrel\ regions where
analytical results do not exist.

In contrast to \nonrel\ shocks, DSA in \rel\ shocks depends critically
on the details of how particles interact with the background magnetic
turbulence. In real shocks, this turbulence must be self generated by
the shock accelerated particles. However, the scattering interactions
responsible for this generation are not known with any reliability and
particle diffusion must be parameterized.
Despite the complexity and unknown nature of the wave-particle
interactions, we believe that much of the essential underlying physics
can be modeled by first assuming that the scattering mean free path
$\lambda$ is proportional to the gyroradius $\rg$ and then defining
two parameters which control the finer scattering details. These
parameters are $\etamfp$ in the relation $\lambda = \etamfp r_g$,
which characterizes the ``strength'' of cross-field scattering, and
the ``fineness'' of pitch-angle scattering $N_g$, where particles
pitch-angle scatter after a fraction of a gyro-period $\deltime =
\tau_g / N_g$ within a maximum angle given by
eqn.~(\ref{eq:Tmax}). 
Spectra converge to a particular form as $N_g$ increases and the
convergent value of $N_g$ is proportional to $\gamZ^2$.

Our results show that particle distributions depend strongly on
$\etamfp$ and $\Tbn$, as well as on $N_g$ for \rel\ shocks when $N_g$
is below the convergent value. Most importantly, we show that $\fofp$
departs significantly from $\fofp \propto p^{-4.23}$ in a wide
\transrel\ regime extending to above $\gamZ = 30$.

In our examples with $\gamZ=2$ we find that the power law is flatter
than $\sigTP =3r/(r-1)$ (Fig.~\ref{fig_trans}) for $\Tbn=0\degg$. That
is, the power-law index $\sigma \simeq 4.12$ versus $\sigTP\simeq
4.29$. As long as $\Tbn=0\degg$, the results are independent of
$\etamfp$. For $\Tbn>0\degg$, the power law steepens as either $\Tbn$
or $\etamfp$ is increased. For our most extreme $\gamZ=2$ case,
$\Tbn=89\degg$ and $\etamfp=6$, yielding $\sigma \simeq 6.5$.

For our more strongly \rel\ examples ($\gamZ \ge 10$), 
with  convergence
values of $N_g$, we obtain the canonical power-law index $\sigma
\simeq 4.23$ for $\Tbn=0\degg$.  As with $\gamZ=2$, the power law
steepens when either $\Tbn$ or $\etamfp$ is increased
(Figs.~\ref{fig_gam10} and \ref{fig_g20_g30}), but the steepening
becomes a weaker function of both $\Tbn$ and $\etamfp$ as $\gamZ$
increases. Our results are consistent with $\sigma \rightarrow 4.23$
independently of $\Tbn$ and $\etamfp$ in the \ultrarel\ limit as
widely reported \citep[e.g.,][]{BedOstrow98,AGKG2001}. We do show,
however, that a wide range of Lorentz factors exists extending up to
$\gamZ \sim 30$ where $\sigma$ differs significantly from $4.23$.

In contrast to the effects of $\Tbn$ and $\etamfp$, which diminish
 with increasing $\gamZ$, the influence of $N_g$ remains as $\gamZ$
 increases. Figs.~\ref{fig_g20_vary_Ng} and \ref{fig_small_Ng} show
 that small $N_g$ (large-angle scattering) produces large features in
 the spectra as individual shock crossings continue to show in $\fofp$
 to high energies. There is no reason to believe this effect will
 diminish with increasing $\gamZ$. 

It is not at all obvious what realistic values of $N_g$ are or if
they are large enough to produce convergence.
Large values of $N_g$ imply that significant power in magnetic
turbulence exists at extremely small length scales and it is possible
that actual plasmas have some lower length-scale limit which may be
larger than needed for convergence.  In this case, the spectrum will
depend on the effective $N_g$ and may be highly variable depending on
the particle shock parameters. Furthermore, the highest momentum
particles require magnetic turbulence with extremely long length
scales for resonance. If this turbulence does not exist, the spectrum
will turn over at some characteristic momentum determined by the
longest magnetic length scale \citep[see][for a discussion of such
effects]{NO2004}.

Some of the most exotic and interesting astrophysical objects are
likely to harbor \rel\ shocks. While it is probable that these shocks
accelerate particles, the spectrum, even in unmodified, \ultrarel\
shocks, depends on the unknown details of the wave-particle
interactions.

\acknowledgments We are grateful to Frank Jones and Matthew Baring for
useful discussions, as well as helpful comments by the referee. This
work was supported, in part, by NASA grant ATP02-0042-0006 and NSF grant
INT-0128883.

\newcommand{\aaDE}[3]{ 19#1, A\&A, #2, #3}
\newcommand{\aatwoDE}[3]{ 20#1, A\&A, #2, #3}
\newcommand{\aatwopress}[1]{ 20#1, A\&A, in press}
\newcommand{\aasupDE}[3]{ 19#1, {\itt A\&AS,} {\bff #2}, #3}
\newcommand{\ajDE}[3]{ 19#1, {\itt AJ,} {\bff #2}, #3}
\newcommand{\anngeophysDE}[3]{ 19#1, {\itt Ann. Geophys.,} {\bff #2}, #3}
\newcommand{\anngeophysicDE}[3]{ 19#1, {\itt Ann. Geophysicae,} {\bff #2}, #3}
\newcommand{\annrevDE}[3]{ 19#1, {\itt Ann. Rev. Astr. Ap.,} {\bff #2}, #3}
\newcommand{\apjDE}[3]{ 19#1, {\itt ApJ,} {\bff #2}, #3}
\newcommand{\apjtwoDE}[3]{ 20#1, {\itt ApJ,} {\bff #2}, #3}
\newcommand{\apjletDE}[3]{ 19#1, {\itt ApJ,} {\bff  #2}, #3}
\newcommand{\apjlettwoDE}[3]{ 20#1, {\itt ApJ,} {\bff  #2}, #3}
\newcommand{\apjpress}{{\itt ApJ,} in press}
\newcommand{\apjletpress}{{\itt ApJ(Letts),} in press}
\newcommand{\apjsDE}[3]{ 19#1, {\itt ApJS,} {\bff #2}, #3}
\newcommand{\apjstwoDE}[3]{ 19#1, {\itt ApJS,} {\bff #2}, #3}
\newcommand{\apjsubDE}[1]{ 19#1, {\itt ApJ}, submitted.}
\newcommand{\apjsubtwoDE}[1]{ 20#1, {\itt ApJ}, submitted.}
\newcommand{\appDE}[3]{ 19#1, {\itt Astropart. Phys.,} {\bff #2}, #3}
\newcommand{\apptwoDE}[3]{ 20#1, {\itt Astropart. Phys.,} {\bff #2}, #3}
\newcommand{\araaDE}[3]{ 19#1, {\itt ARA\&A,} {\bff #2},
   #3}
\newcommand{\assDE}[3]{ 19#1, {\itt Astr. Sp. Sci.,} {\bff #2}, #3}
\newcommand{\grlDE}[3]{ 19#1, {\itt G.R.L., } {\bff #2}, #3} 
\newcommand{\icrcplovdiv}[2]{ 1977, in {\itt Proc. 15th ICRC (Plovdiv)},
   {\bff #1}, #2}
\newcommand{\icrcsaltlake}[2]{ 1999, {\itt Proc. 26th Int. Cosmic Ray Conf.
    (Salt Lake City),} {\bff #1}, #2}
\newcommand{\icrcsaltlakepress}[2]{ 19#1, {\itt Proc. 26th Int. Cosmic Ray Conf.
    (Salt Lake City),} paper #2}
\newcommand{\icrchamburg}[2]{ 2001, {\itt Proc. 27th Int. Cosmic Ray Conf.
    (Hamburg),} {\bff #1}, #2}
\newcommand{\JETPDE}[3]{ 19#1, {\itt JETP, } {\bff #2}, #3}
\newcommand{\jgrDE}[3]{ 19#1, {\itt J.G.R., } {\bff #2}, #3}
\newcommand{\mnrasDE}[3]{ 19#1, {\itt MNRAS,} {\bff #2}, #3}
\newcommand{\mnrastwoDE}[3]{ 20#1, {\itt MNRAS,} {\bff #2}, #3}
\newcommand{\mnraspress}[1]{ 20#1, {\itt MNRAS,} in press}
\newcommand{\natureDE}[3]{ 19#1, {\itt Nature,} {\bff #2}, #3}
\newcommand{\naturetwoDE}[3]{ 20#1, {\itt Nature,} {\bff #2}, #3}
\newcommand{\nucphysA}[3]{#1, {\itt Nuclear Phys. A,} {\bff #2}, #3}
\newcommand{\pfDE}[3]{ 19#1, {\itt Phys. Fluids,} {\bff #2}, #3}
\newcommand{\phyreptsDE}[3]{ 19#1, {\itt Phys. Repts.,} {\bff #2}, #3}
\newcommand{\physrevEDE}[3]{ 19#1, {\itt Phys. Rev. E,} {\bff #2}, #3}
\newcommand{\prlDE}[3]{ 19#1, {\itt Phys. Rev. Letts,} {\bff #2}, #3}
\newcommand{\revgeospphyDE}[3]{ 19#1, {\itt Rev. Geophys and Sp. Phys.,}
   {\bff #2}, #3}
\newcommand{\rppDE}[3]{ 19#1, {\itt Rep. Prog. Phys.,} {\bff #2}, #3}
\newcommand{\rpptwoDE}[3]{ 20#1, {\itt Rep. Prog. Phys.,} {\bff #2}, #3}
\newcommand{\ssrDE}[3]{ 19#1, {\itt Space Sci. Rev.,} {\bff #2}, #3}
\newcommand{\ssrtwoDE}[3]{ 20#1, {\itt Space Sci. Rev.,} {\bff #2}, #3}
\newcommand{\scienceDE}[3]{ 19#1, {\itt Science,} {\bff #2}, #3} 
\newcommand{\spDE}[3]{ 19#1, {\itt Solar Phys.,} {\bff #2}, #3} 
\newcommand{\spuDE}[3]{ 19#1, {\itt Sov. Phys. Usp.,} {\bff #2}, #3}

\end{document}